\documentclass[twocolumn]{aa}
\usepackage{graphicx}
\usepackage{txfonts}
\begin{document}
   \title{{\em XMM-Newton} observation of the ULIRG NGC~6240}

   \subtitle{The physical nature of the complex Fe K line emission}

    \author{Th. Boller
        \inst{1}
        \and
        R. Keil
        \inst{1}
        \and
        G. Hasinger
        \inst{1}
        \and
        E. Costantini
        \inst{1}
        \and
        R. Fujimoto
        \inst{2}
        \and
        N. Anabuki
        \inst{2}
        \and
        I. Lehmann
        \inst{1}
        \and
        L. Gallo
        \inst{1}
          }

   \offprints{Th. Boller: bol@mpe.mpg.de}

   \institute{Max-Planck-Institut f\"ur extraterrestrische Physik, Postfach 1312, 85741 Garching, Germany
         \and
    Institute of Space and Astronautical Science, 3-1-1 Yoshinodai, Sagamihara, Kanagawa 229-8510, Japan
             }

   \date{Received -- ; accepted -- }

   \abstract{
We report on an {\em XMM-Newton} observation of the ultraluminous
infrared galaxy NGC~6240. The 0.3--10 keV spectrum can be
successfully modelled with: (i) three collisionally ionized plasma
components with temperatures of about 0.7, 1.4, and 5.5 keV; (ii)
a highly absorbed direct power-law component; and (iii) a neutral
Fe K$\rm \alpha$ and K$\rm \beta$ line. We detect a significant neutral column
density gradient which is correlated  with the temperature of the three
plasma components. Combining the {\em XMM-Newton} spectral model
with the high spatial resolution {\em Chandra} image we find that the
temperatures and the column densities increase towards the center.

With high significance, the Fe K line complex is resolved
into three distinct narrow lines: (i) the neutral Fe K$\rm
\alpha$ line at 6.4 keV; (ii) an ionized line at about 6.7 keV;
and (iii) a higher ionized line at 7.0 keV (a blend of the Fe XXVI and the Fe K$\rm \beta$ line).
While the neutral Fe K line is most probably due to reflection from optically thick
material, the  Fe XXV and Fe XXVI emission arises from  the highest temperature
ionized plasma component.

We have compared the plasma parameters of the ultraluminous infrared galaxy NGC 6240 with those found in the local starburst galaxy
NGC 253.  We find a striking similarity in the plasma temperatures and column density gradients, suggesting a
similar underlying physical process at work in both galaxies.

   }

   \maketitle
%

\section{Introduction}

The  Infrared Astronomical Satellite (IRAS) mission (Neugebauer et al. 1984)
resulted in the discovery of a large number of galaxies in the local universe
($\rm z < 0.3$) with extraordinarily large infrared luminosities 
($\rm L_{IR}\ge 10^{12}\,L_{\odot}$ for $\rm H_0 = 75\,km\,s^{-1}\,Mpc^{-1}$; see Sanders \& Mirabel 1996 for a
review).  The number density of these Ultraluminous InfraRed Galaxies (ULIRGs)
exceeds that of optically selected Seyfert galaxies and QSOs with comparable
bolometric luminosities (Soifer et al. 1987; Sanders et al. 1988a \& 1988b; Sanders \&
Mirabel 1996) by a factor of $\rm \sim 1.5-2$ (Sanders et al. 1999).
Some of the most fundamental issues regarding ULIRGs are: the
nature of the power source (strong starburst and/or a
dust-enshrouded AGN); and the relative contribution
to their bolometric luminosity.
The capability of X-rays to penetrate high absorbing material
makes X-ray observations essential in attempting to understand
the physical processes at work in ULIRGs.
A considerable portion of ULIRGs have been found to contain a hard X-ray source,
highly absorbed by a molecular torus, which indicates the presence of
a hidden AGN
(Mitsuda 1995; Brandt et al. 1997; Kii et al. 1997; Vignati et al. 1999; Braito et al. 2003, Ptak et al. 2003, Franceschini et al. 2003).

NGC~6240 (IRAS 16504+0228) is perhaps one of the most interesting
and best studied objects in this class. Although it is slightly less
luminous ($\rm L_{FIR} \simeq 6 \cdot 10^{11}\,L_{\odot}$ for
$\rm H_0 = 75\,km\,s^{-1}\,Mpc^{-1}$; Genzel et al. 1998) than the
typical IR luminosity used to define a ULIRG, it exhibits all of the
ULIRG characteristics, and it is thus studied as a prototype of this
class of sources. Optical/IR observations
show a complex system with a disturbed morphology consisting of
two gravitationally interacting nuclei (Fosbury \& Wall 1979;
Fried \& Schulz 1983). 
An $ISO$ observation of NGC~6240 (Genzel et al.
1998) favors a starburst-dominated power source, in which the high
infrared emission arises from warm, absorbing dust, surrounding
the inner parts of the galaxy.  Early X-ray observations of
NGC~6240 with $ROSAT$ (Schulz et al. 1998) revealed an exceptionally high X-ray
luminosity ($\rm L_{0.1-2.0\,keV} \sim 10^{10}\,L_{\odot}$), which
further confirmed the importance of NGC~6240 in understanding the
AGN-starburst connection. Therefore, considerable effort has been
made to analyze the spectral and spatial properties using
different X-ray satellites. Moreover, NGC 6240 can be considered
as the prototype objects of highly absorbed AGN in the local
universe (Hasinger 2003).

First {\em ROSAT}  observations showed that a hot, diffuse gas model in combination with a
power-law component can give a good
description of the $\rm (0.1 - 2.4)\,keV$ energy band spectrum (Schulz
et al. 1998). The power-law component was interpreted as evidence  for a hidden AGN in NGC 6240.
A luminous extended soft X-ray component was reported by Komossa,
Schulz \& Greiner (1998) using {\em ROSAT} HRI observations. This
emission should be responsible for at least 60\% of the $\rm 0.1 -
2.4\,keV$ luminosity (within a region of $\rm 5''$ radius from the
center). The combination of the spatial extent and the Raymond-Smith-like
thermal spectrum led to the idea that the origin of the soft X-rays
were strong thermal processes indicative of a powerful, X-ray luminous
starburst.

According to data from {\em ROSAT} and {\em ASCA},
Iwasawa \& Comastri (1998) reported the detection of a highly absorbed hard X-ray
source in NGC~6240. They pointed out that the soft X-ray spectrum ($\rm \sim
0.5 - 2.0\,keV$) could be described by a superposition of two
optically thin thermal plasmas with different temperatures ($\rm kT_1 \simeq
0.2-0.6\,keV$ and $\rm kT_2 \simeq 1\,keV$) and an $\rm N_H$
value roughly an order of magnitude higher than the Galactic value of
$\rm N_{H,gal} = 5.8 \cdot 10^{20}$ atoms $\rm cm^{-2}$.
At higher energies ($\rm \sim 3 - 10\,keV$) {\em ASCA} observed an extremely
hard continuum slope accompanied by the presence of a strong Fe K$\rm \alpha$
complex (Mitsuda 1995; Iwasawa \& Comastri 1998). This can be interpreted in
terms of a  luminous AGN completely hidden by Compton-thick
material, whose emitted light is reflected into the line-of-sight (e.g. George,
Nandra \& Fabian 1990). In addition, the Fe complex could be well explained by
a superposition of two Gaussian lines at $\rm E_1 = (6.44 \pm 0.04)\,keV$ and
$\rm E_2 = (6.87 \pm 0.05)\,keV$, this latter was interpreted as a strong
hint for a highly ionized reflecting material.

Vignati et al. (1999) and Ikebe et al. (2001) investigated NGC~6240 by using
data from {\em BeppoSAX} and {\em RXTE}, respectively. Both authors assert the
existence of an AGN component directly related to a Compton thick absorber
(with $\rm N_H \simeq 2 \cdot 10^{24}$ atoms $\rm cm^{-2}$).

Recent {\em Chandra} observations have spatially resolved the emission
from the two nuclei and the surrounding starbursts at X-rays. The presence
of the neutral Fe K$\rm \alpha$ lines and  the hard power-law-like
X-ray emission from both nuclei was used to
infer that both nuclei of the galaxy harbours an AGN (Komossa et al. 2003).
In addition, they found residua in their spectral fits at energies above the neutral
Fe K line, which they interpreted as possible emission from H-like Fe from the
nuclear regions.

This paper contains the analysis of the X-ray data of NGC~6240 taken with {\em
XMM-Newton} (Jansen 1999). The data analysis and details of the
{\em XMM-Newton} observations are described in Section 2. Section 3
contains the spectral fitting results. A comparison with the {\em Chandra} observations is given in Section 4.
The summary of the {\em XMM-Newton} results on NGC 6240 is
given in Section 5. We use a Hubble constant of $\rm H_0 = 50\ km\ s^{-1}\  Mpc^{-1}$ and $\rm q_0 = 0$ throughout.

\section{Technical details: {\em XMM-Newton} observation and data processing}

NGC~6240 was observed with {\em XMM-Newton} for the first time during orbit 0144
on the $\rm 22^{nd}$ of September, 2000, with all instruments functioning except for the RGS.
The EPIC pn camera was operated in the full-frame mode and the EPIC MOS cameras
were operated in the large-window mode. Each of the detectors used the medium filter.
The exposure times were 25150 s and 24349 s for
the pn and MOS, respectively.
The second observation, with a shorter exposure time of about 18 ks,
was performed during orbit 0413 ($\rm 12^{th}\,of\,March, 2002$),
with all instruments working in the same mode and with the same filters as in the first exposure. In addition, the
RGS was functional and operated in the Spectro+Q mode, however the exposure time was too low to collect
sufficient signal (raw exposure times: 18917~s for each of the RGS devices).

The MOS1 and MOS2 data sets from each observation were merged for this analysis.
Single and double events have been selected and the resulting PHA file was grouped 
with a minimum of 20 counts per bin.
After correcting the MOS data for background flaring events, the GTI corrected exposure times for MOS1 and MOS2 were
23381~s and 23601~s (first observation -- orbit 0144)  and 11318~s and
11601~s (second observation -- orbit 0413).
This gives us a total of about 35
ks worth of exposure time which is used in the analysis.
Examining each observation separately and comparing them to the
merged data set, we found no discrepancies in the spectral fitting
parameters within the data uncertainties, justifying the merging
process. In addition, the EPIC pn data sets were merged in a similar fashion (orbit 0144: 16255~s, orbit 0413: 9750~s).

The EPIC MOS energy resolution (FWHM) is slightly better than that of the EPIC pn. Comparing Fig. 25 and 26 of the XMM-Newton User's Handbook (v2.1; Document
number: XMM-PS-GM 14\footnote{{\tt
ftp://xmm.vilspa.esa.es/pub/AO3/XMM\_UHB.pdf}}) one gets an energy
resolution for the EPIC MOS cameras of about 140 eV between 6 and 7
keV. For the EPIC pn camera the energy resolution ranges between 160
and 240 eV, but it is better for double events (upper curve of
Fig. 26) and decreases for single events when considering on--axis
sources. As we have used single and double events, the EPIC pn energy
resolution ranges between 160 and 240 eV (the exact value depends on the ratio between single and double events which is presently not 
available). In order to utilize as many photons as possible we have
concentrated on the EPIC MOS data for subsequent spectral analyses (about twice as many
photons than in the EPIC pn data). With the combined MOS1/2
photons from the two observations we collect in the 0.3--10 keV energy range
about $\rm 1.0 \cdot 10^{-3}$ $\rm  photons\ cm^{-2}\ s^{-1}\ keV^{-1}$ (compared to
 $\rm 5 \cdot 10^{-4} \rm  photons\ cm^{-2}\ s^{-1}\ keV^{-1}$ for EPIC pn). Nevertheless
we use the EPIC pn data to independently check the EPIC MOS
results for consistency.

The data were processed using the standard
procedures of the {\em XMM-Newton} Science Analysis System
(XMM-SAS\footnote{{\tt http://xmm.vilspa.esa.es/sas/ }}) version
20020413\_2031-5.3.0 with default parameters for the EPIC chains. These
event lists were calibrated with the latest available calibration
files\footnote{{\tt http://xmm.vilspa.esa.es/ccf/}}.

 Source and background counts were extracted from circular regions
with radius of $70''$.
The background photons were extracted from a source free region with the same radius.
For the X-ray spectral analysis we have used XSPEC version 11.2 (Arnaud 1996) as well as
FTOOLS version 5.2.  The quoted errors on the derived best-fitting model parameters
correspond to a 90\% confidence level.

\begin{figure}
\includegraphics[width=6.4cm,angle=-90,clip=]{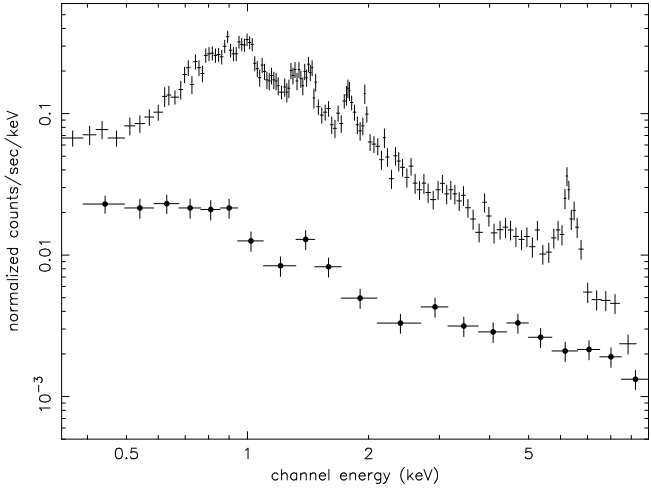}
\caption{The $\rm (0.3 - 10)\; keV$ spectrum of the combined source and background counts (upper curve) and the background counts (lower curve with the black dots) for the EPIC MOS1/2 data. The spectrum is not background dominated below 10 keV and the Fe K line complex can be investigated with a high signal-to-noise ratio.
}
\end{figure}

\section{Spectral analysis}

The combined MOS1/2 spectrum of NGC 6240 is not background
dominated below 10 keV (c.f. Fig. 1); therefore, the energy range between $\rm (0.3 - 10)\,keV$ is analyzed with relatively good signal-to-noise.

\subsection{The soft energy spectrum}

\subsubsection{Line identification}
First, we searched for emission lines in the soft
spectral region ($\rm 0.3 - 3.5\,keV$). Figure 2 shows the EPIC MOS1/2 source
spectrum and several identified emission lines.  For illustration purposes we
have included a 1-keV-model spectrum of a diffuse
collisionally-ionized plasma, calculated
using the APEC algorithm (Astrophysical Plasma Emission
Code\footnote{{\tt http://hea-www.harvard.edu/APEC/}}; Smith et al.
2001) with variable abundances as provided by the vAPEC model in
XSPEC (solid line). 
The abundances of O, Ne, Mg, Si, S, and Ar were fixed to 10 times solar and
the abundance of Fe was fixed to 0.01 times solar (the others were fixed
to solar abundances).  Pairs of emission lines, which are in
different ionization states, can be resolved in the EPIC MOS1/2
spectrum. Moreover, higher energy emission line such as S XV and
Ar XVII are also detected. The ratio of the emission line pairs
and the emission lines at higher energies enable us to determine
how many thermal plasma components are required to fit the soft energy spectrum.

\begin{figure}
\begin{minipage}{8.3cm}
\includegraphics[width=6.0cm,angle=-90,clip=]{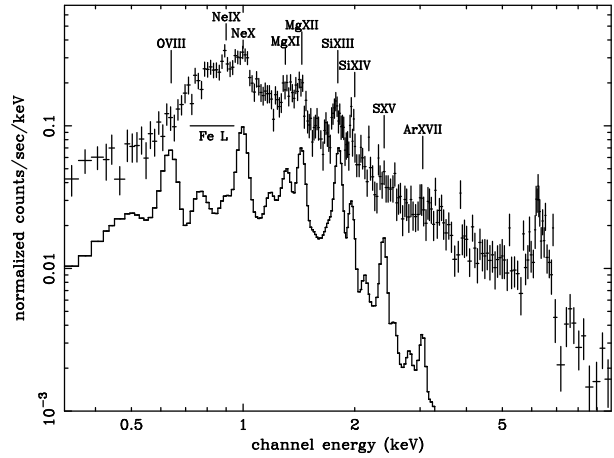}
\caption{EPIC MOS1/2 spectrum of NGC 6240  (upper curve). For illustration
purposes a vAPEC model with a
temperature of 1 keV has been included (lower line). }
\end{minipage}
\end{figure}

\begin{figure}
\includegraphics[width=6.0cm,angle=-90,clip=]{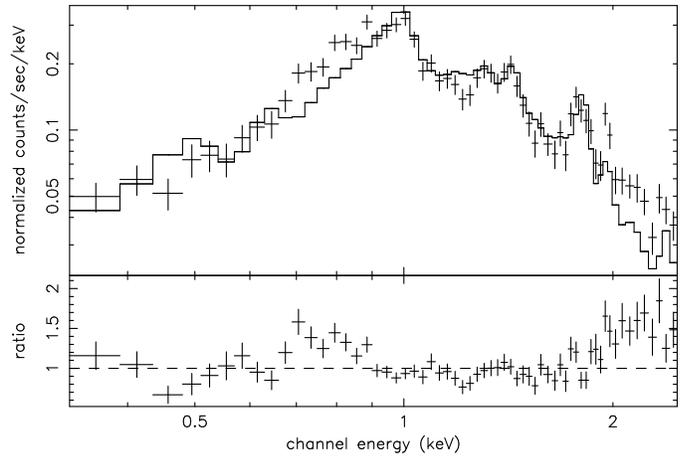}
\caption{Fitting only one vAPEC model to the EPIC MOS1/2 data in the (0.3--2.5)~keV range,
where O, Ne, Mg, Si, S; the other
$\alpha$ elements; and Fe--Ni abundances have been left free.
The resulting fit is statistically unacceptable.} \vskip 0.2cm
\end{figure}

\subsubsection{The presence of two distinct plasmas}

Applying only one vAPEC model to the EPIC MOS1/2 data with thawed abundance parameters for O, Ne,
Mg, Si, S, the $\alpha$ elements, and Fe--Ni,  did not result in an acceptable fit.
As can be seen in Fig. 3, significant residua remain
in the Fe~L emission line
region around 0.8 keV, as well as in the $\rm (1.0 - 1.5)\,keV$ range  ($\rm \chi_r^2 = 1.5$).
The observed Fe~L bump, the flux ratio between two emission lines in different
ionization states and the fit residuals all suggest the presence of a second thermal component.

We have added a second vAPEC component to the first model, which
resulted in an acceptable fit  ($\rm \chi_r^2 = 1.0$). The
significance of this new component was tested with the F-test. The
second component is required with a significance greater than 
5$\rm\sigma$. The pn data explicitly confirm the EPIC MOS1/2
results.  The best fit parameters of the two vAPEC
components are reported in Table 1.

\subsection{The Fe K line complex}

\begin{figure}
\includegraphics[angle=-90,width=8.3cm,clip=]{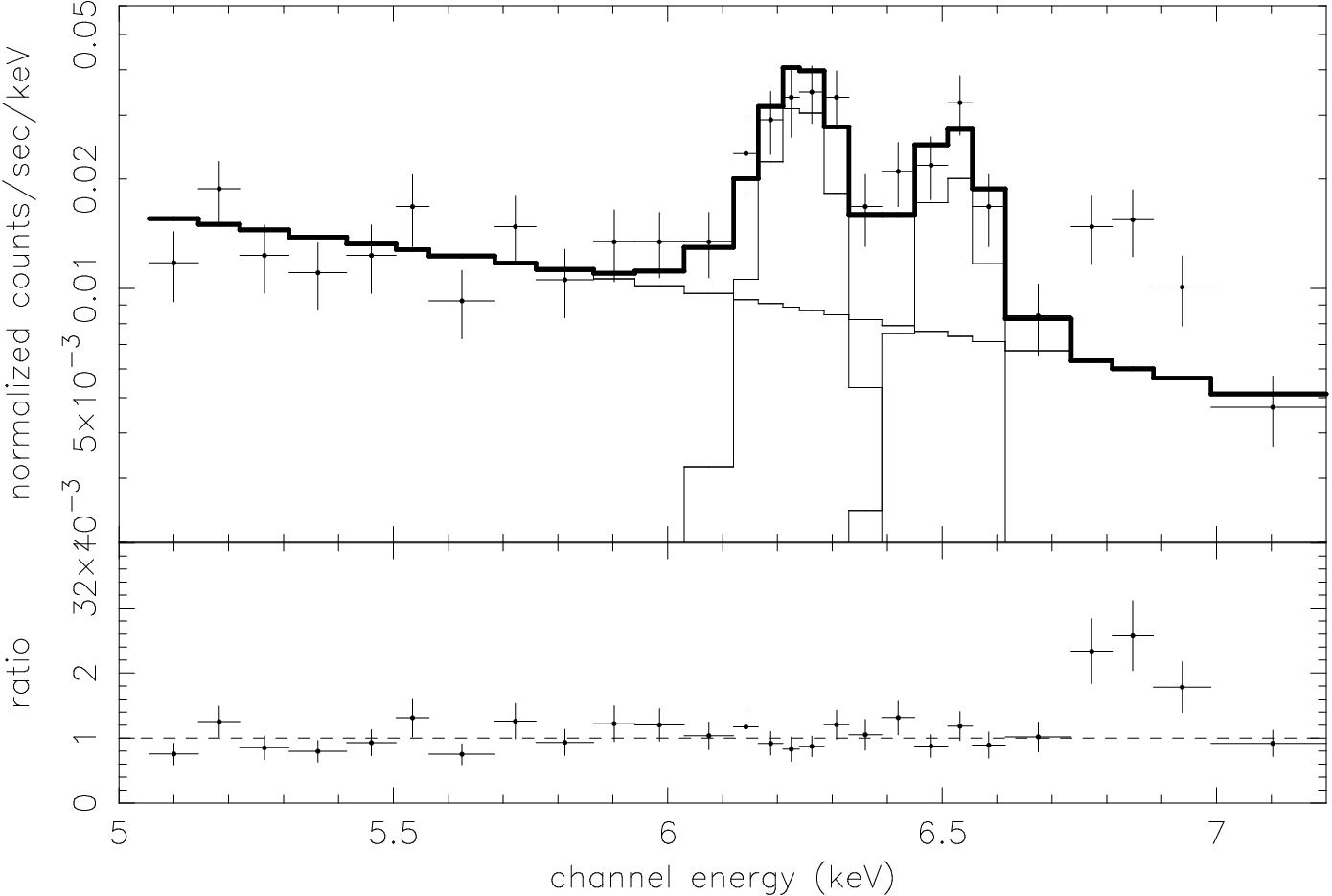}
\includegraphics[angle=-90,width=8.3cm,clip=]{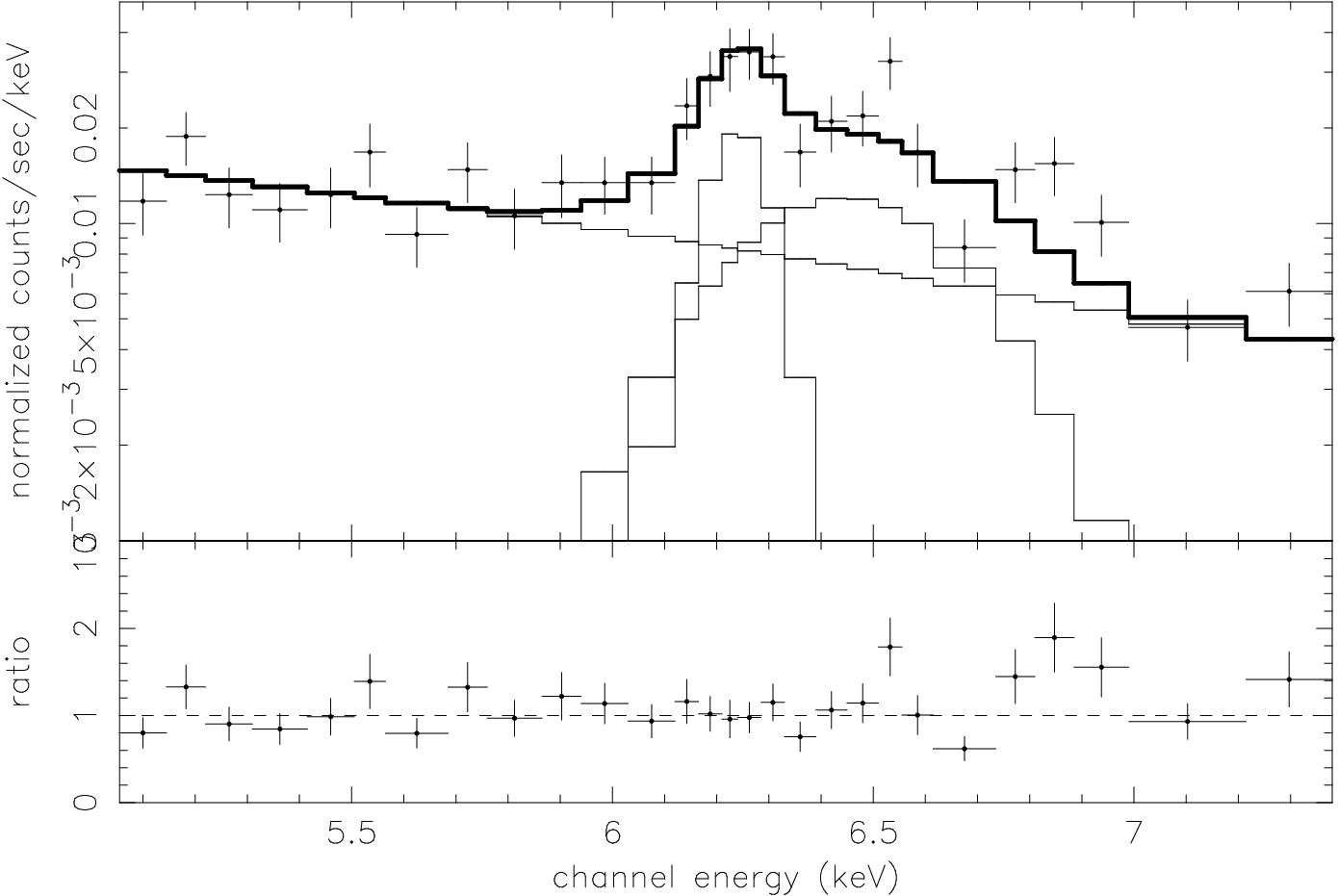}
\includegraphics[angle=-90,width=8.3cm,clip=]{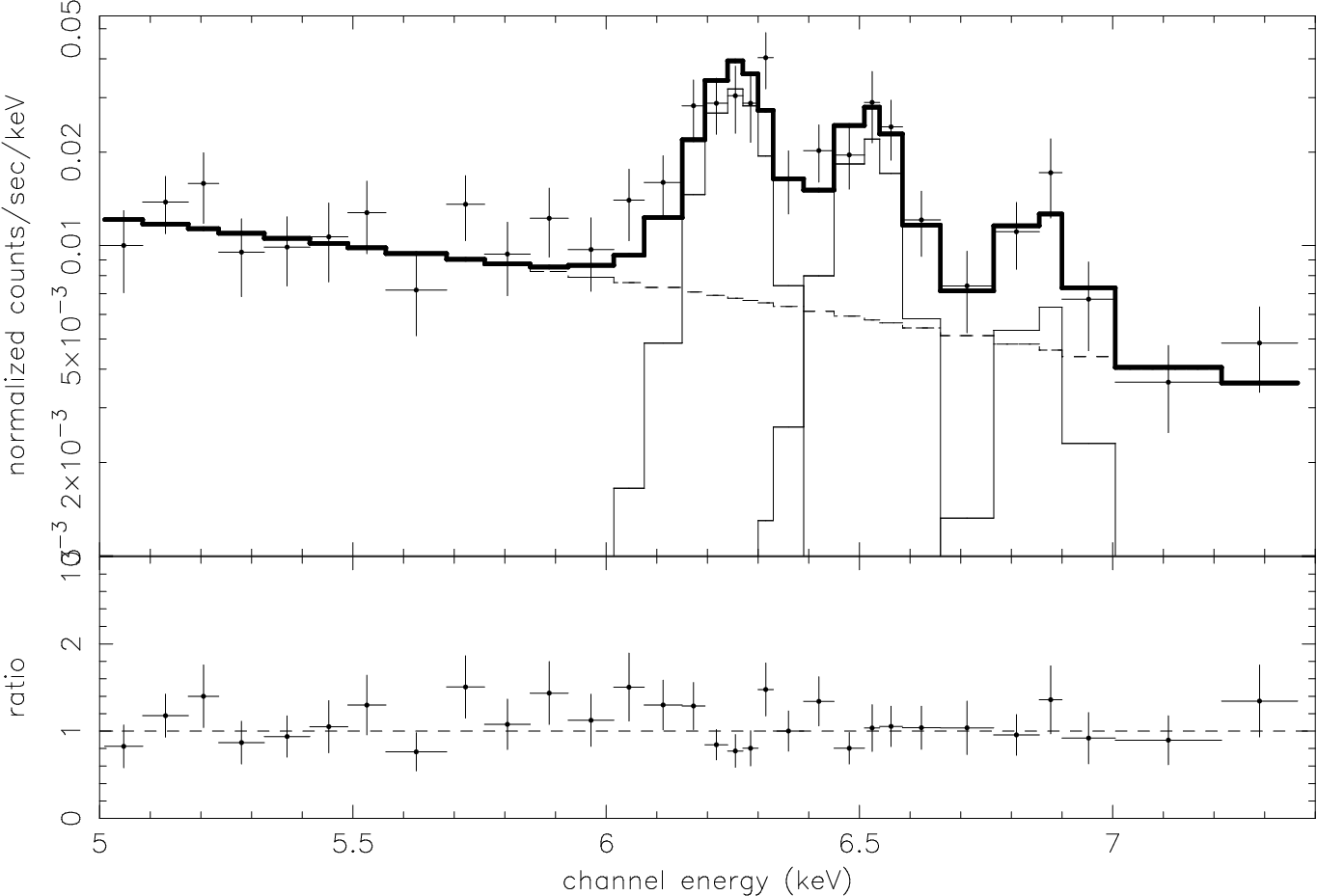}
\caption{{\bf Upper panel}: Spectral fit to the {\em XMM-Newton} combined MOS1/2 data of NGC 6240 in the Fe K line range.
Fitting the Fe line complex with only two narrow lines and a  power-law continuum results in a statistically unacceptable fit.\newline {\bf Middle panel}:
 Fitting the Fe line complex with a narrow neutral line and one broad line also result in a statistically unacceptable fit. \newline {\bf Bottom panel}:
 An acceptable fit is obtained when fitting three Gaussian lines above the power-law continuum. The centroid line energies are at
(6.41 $\pm$ 0.02) keV,
(6.68 $\pm$ 0.02) keV and
(7.01 $\pm$ 0.04) keV. With high statistical significance, the {\em XMM-Newton} data  reveal for the first time  the presence of three  narrow Fe K lines.}
\end{figure}

The high throughput of {\em XMM-Newton} allows for  precise measurements of
the Fe K line complex compared to previous X-ray missions. In this  Section
we will demonstrate with high significance that the Fe K line complex is resolved into three distinct lines.
The line energies are given in the rest frame of the source.

\subsubsection {{\em XMM-Newton} detection of three distinct Fe K emission lines}

As a simple first approach the $\rm 5 - 7.2\,keV$ energy range has
been fitted with a power-law model and two Gaussian lines (c.f.
Fig. 4, upper panel). The centroid line energies are found at (6.5
$\pm$ 0.3) keV and (6.68 $\pm$ 0.02) keV. Strong residua remain at
around 6.9 keV and the fit is statistically unacceptable ($\rm
\chi^2$ = 30 for 22 d.o.f.). We can exclude the possibility that
the Fe line complex can be modelled by a narrow unresolved 6.4 keV
line and a blend of  higher ionized lines (Fig. 4, middle panel;
$\rm \chi^2$  =  30 for 20 d.o.f.).

The strong residuals which remain at $\rm 6.8 - 6.9\,keV$ clearly 
are reduced by using a power-law plus three narrow unresolved Gaussian lines
($\rm \chi^2$ = 15 for 19 d.o.f., see Fig. 4, bottom panel).
The centroid line energies and
corresponding 90 per cent coincidence ranges are (6.41 $\pm$ 0.02)
keV, (6.68 $\pm$ 0.02) keV and (7.01 $\pm$ 0.04) keV.
With high statistical significance, the {\em XMM-Newton} data  reveal for the first time  the presence of three  narrow Fe K lines.

\subsubsection{The possible physical origin of the ionized Fe K lines}

\begin{figure}
\includegraphics[angle=-90,width=8.3cm,clip=]{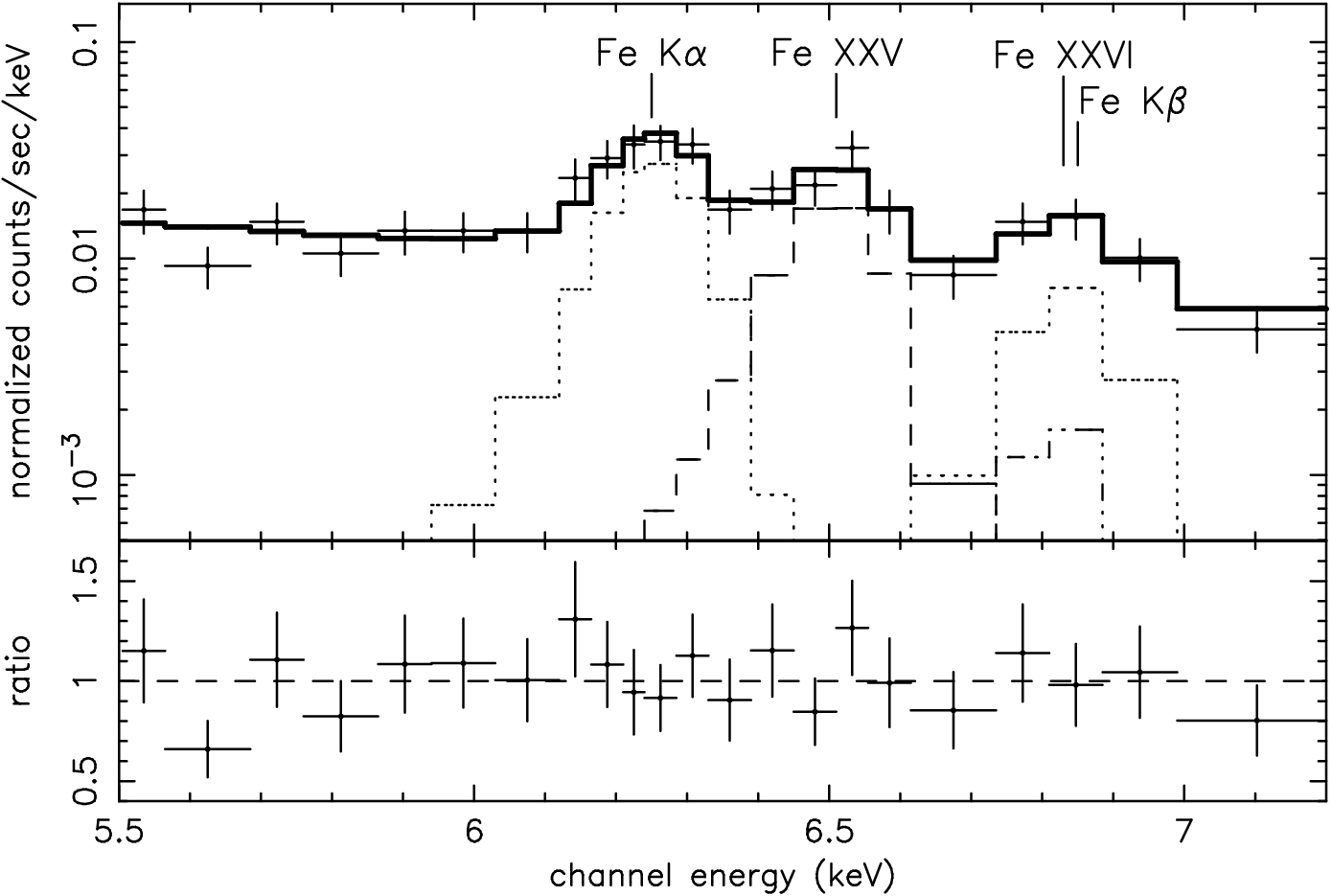}
\caption{Spectral fit to the {\em XMM-Newton} MOS1/2 data of NGC 6240 in
the Fe K line range (with the identifications of the Fe lines). Emission from a collisionally ionized plasma has been
assumed to model the ionized Fe K line emission. The neutral 6.4 Fe~K$\rm \alpha$ line has been
modelled by a simple Gaussian line. We have also included the emission from the Fe K$\rm \beta$ line
where the ratio has been fixed to 1/8.8 according to atomic physics. The Fe XXV and Fe XXVI lines originate from the hottest plasma component.}
\includegraphics[angle=-90,width=8.3cm,clip=]{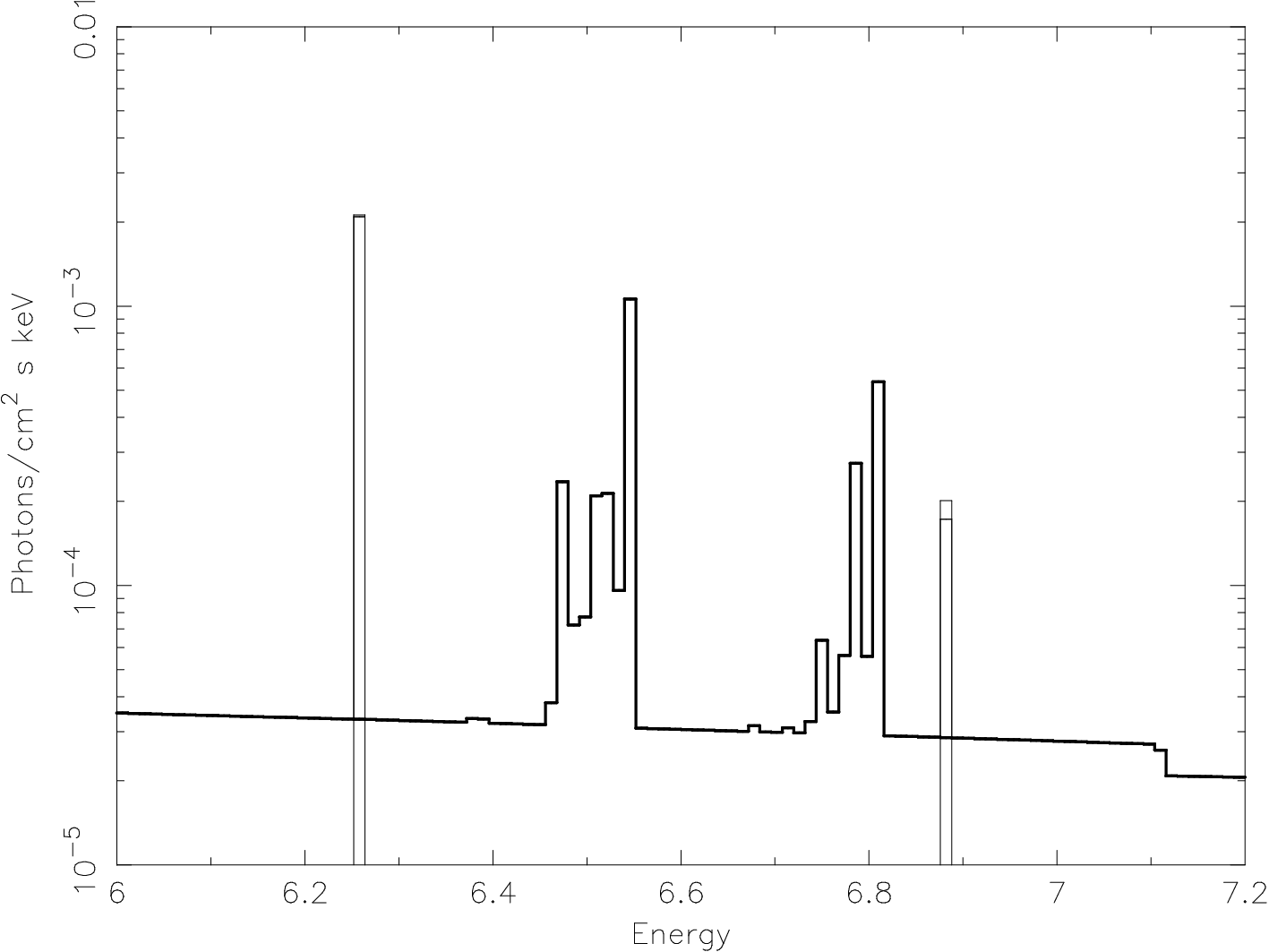}
\caption{Spectral model for the Fe K line complex. Two unresolved Fe lines (K$\rm \alpha$ and K$\rm \beta$)
and emission from a collisionally ionized plasma (APEC model in
XSPEC) can explain the emission from the Fe K line complex in NGC 6240 (for line identifications see Fig. 5).}
 \end{figure}

One plausible model to explain the presence of ionized Fe K emission lines arises
from the presence of a collisionally
ionized plasma (using the APEC model in XSPEC). The plasma with a
temperature of about 5.5 keV produces the Fe XXV (at 6.68 keV) and Fe XXVI (6.93 keV)
emission lines (c.f. also the discussion of the broad band spectrum in Section 3.3).
 In the spectral fit for the Fe line complex we
have also included the emission from the Fe K$\rm \alpha$ line at 6.4
keV and the Fe K$\rm \beta$ line at 7.058 keV, each modelled with
a Gaussian line. The normalization in the spectral fit of the line
intensity of the Fe K$\rm \beta$ line was set to 1/8.8 of the Fe
K$\rm \alpha$ line intensity, as expected from atomic physics. Fig. 5
shows the EPIC MOS1/2 spectral fitting results for the Fe line
complex in NGC 6240 ($\rm \chi_{r}^{2}$ = 1.1). The model
components are displayed in Fig. 6.

In addition, we tested other possibilities to describe the origin of the ionized line emission in NGC 6240.
First, we modelled the Fe K line complex by He- and H-like emission from
the accretion disk. Using the Ballantyne et al.  (2001) model available within XSPEC
we are unable to find a statistically acceptable fit.
This is  expected as {\em BeppoSAX} observations show that the direct AGN component is heavily
absorbed and only dominates the spectrum above 8 keV (c.f. Fig. 3
of Vignati et al. 1999). Therefore, the putative He- and H-like
emission from the accretion disk should remain undetectable.

Secondly, we substituted the APEC model by an absorbed power-law component plus three Gaussian lines.
The fit to the $\rm (0.3 - 10)\,keV$ spectrum was statistically unacceptable ($\rm
\chi_{r}^{2}$ = 1.8) and  is unable to explain the broad-band spectral
energy distribution, nor the presence of ionized Fe K lines.

Thirdly, fitting the broad-band spectrum with only two APEC
components was
statistically unacceptable. A third APEC component is required by the data. The two lower temperature APEC
components have already been discussed in section 3.1.2.

Colbert et al. (2002) have modelled the Fe K features in NGC 1068 from both
optically thick and optically thin ionized reflection emission.  The optically thick
case was modelled using the PEXRIV model in XSPEC, the optically thin case
by a power-law plus edge model. The authors find that the Compton reflection
component provided by the optically thick model does give a better fit than the optically-thin
reflection model. We have modelled the {\em XMM-Newton} spectrum by substituting the APEC model
with the PEXRIV model (and by adding two Gaussian lines to model the ionized Fe K lines). This model
does not provide an acceptable fit ($\rm \chi_{r}^{2}$ = 4.6). The same holds for the power-law
plus edge model for an optically thin warm reflector. The {\em XMM-Newton} data do not require, within
the available statistics, an additional ionized edge.

The collisionally ionized plasma provides the best spectral fitting results
and appears the most likely explanation for these {\em XMM-Newton} observations (see Table 1 for the best fitting parameters).

\subsection{The broad-band X-ray spectrum}

\begin{figure}
\includegraphics[width=6.0cm,angle=-90,clip=]{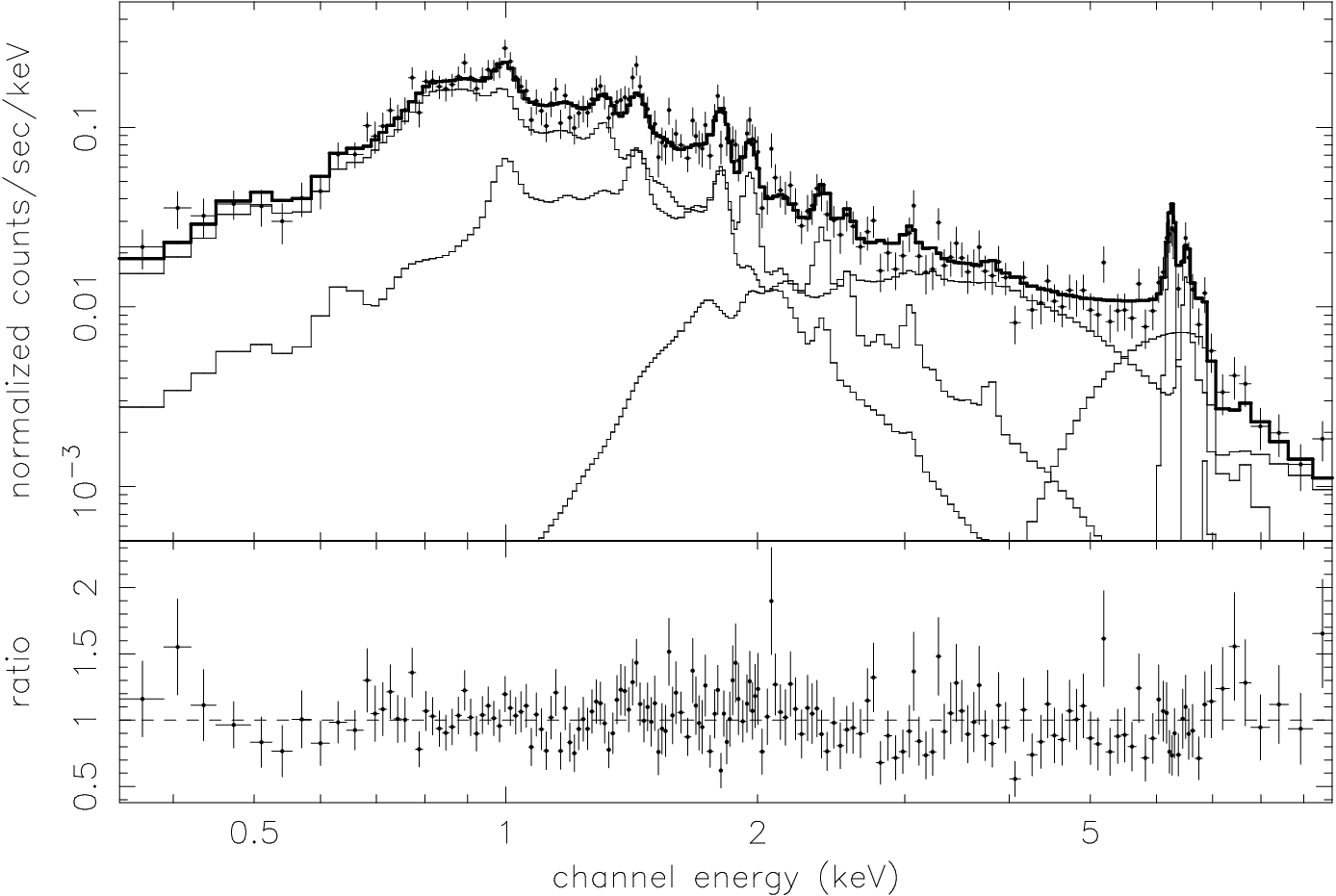}
\caption{Spectral fit to the merged MOS1/2  data of NGC
6240 in the 0.3--10 keV energy range. In the soft energy band, a statistically acceptable
fit is obtained by using a hot and cool plasma component, where
the $\rm \alpha$ and  Fe-Ni abundances have been left free in the
fit. The emission from a third hot plasma component
 is also required to reproduce the line ratio of the ionized line
emission of the Fe K complex. A narrow unresolved Fe K$\rm \alpha$ and Fe K$\rm \beta$ line, with a relative ratio of  8.8 : 1, as expected
from atomic physics, has been added to model.  A highly absorbed ($\rm N_H = (1.0 \pm
0.3) \cdot 10^{24}\ cm^{-2}$) power-law component with an photon
index of 1.8 significantly improves the fit above about 7 keV. The
fit to the 0.3--10 keV band is statistically acceptable with the
above mentioned components ($\rm \chi^2$ = 0.93 for 190 d.o.f). }
\end{figure}

\begin{table*}
\begin{center}
\begin{tabular}{lllllllll}
\hline
(1)                                                             &       (2)                             &      (3)                             &      (4)                                 &       (5)                            &         (6)                                      &                  (7)                          &                  (8)             &        (9)             \\
\hline \hline
                                                                  &    vAPEC(1)                        &   vAPEC(2)                        &    APEC(3)                           & power-law                     &     Fe K$\rm \alpha$            &                 Fe XXV                   &             Fe XXVI                  &    Fe K$\rm \beta$          \\
\hline
$\rm N_H(fit)$ $[\rm 10^{22}\ cm^{-2}]$   & 0.20 $\pm$ 0.03            & 0.4 $\pm$ 0.03           & 4.1 $\pm$ 1.3                   &  100  $\pm$ 20             &                      -                              &                   -                            &                     -               &          -           \\
kT [keV]                                                     & 0.66 $\pm$ 0.03           &  1.4 $\pm$ 0.2               & 5.5 $\pm$ 1.5                 &         -                               &                      -                            &                 -                               &                   -                     &        -                \\
Norm $^a$                                   & $\rm 4.8 \pm 1.2$                &  $2.5 \pm 1.5$                & $3.7 \pm 1.3$                 &  $6.0 \pm 3.0$               &     $2.6 \pm 0.8$              &     $1.8 \pm 0.7$          &      $0.9 \pm 0.7$   &  $0.3 \pm 0.1$              \\
 f [$\rm 10^{-12}\,erg\,cm^{-2}\,s^{-1}$]                    & $1.5 \pm 0.4$   & $0.7 \pm 0.4$  &  $1.4 \pm 0.5$   &  $40 \pm 20$ &  $0.25 \pm 0.08$        &  $0.19 \pm 0.07$      &     $0.10 \pm 0.08$     & $0.03 \pm 0.01$                                 \\
L [$\rm 10^{42}\,erg\,s^{-1}$]                                 &  $3.9 \pm 1.0$   &  $1.9 \pm 1.1$  & $3.6 \pm 1.3$   &  $100 \pm 50$ &     $0.60 \pm 0.20$         &    $0.50 \pm 0.20$     &     $0.30 \pm 0.20$  & $0.08 \pm 0.02$                 \\
Equivalent width [eV] & - & - & - & - &  $\rm 300 \pm 100$ &  $\rm 220 \pm 90$ & $\rm 120 \pm 90$ &  $\rm 50 \pm 20$ \\
\end{tabular}
\caption{Spectral fit components to the (0.3--10) keV energy band. The individual
columns 2 to 8 refer to the spectral components applied in the fit. The rows specify the measured values. The broad-band spectral energy distribution
measured with these components is shown in Fig. 7. The foreground absorption has been fixed to the Galactic value.
$^a$ NOTE: For the first three components the normalization
is given in units of [$\rm 10^{-14} / (4 \pi (D_A \cdot (1+z))^2) \int n_e\ \cdot  n_H\ \cdot dV$],
where $\rm D_A$ is
          the angular size distance to the source (cm), $\rm n_e$ is the electron
          density ($\rm cm^{-3})$, and $\rm n_H$ is the hydrogen density ($\rm cm^{-3}$). For the last five components the
	  normalization is given in units of [$\rm 10^{-5}\,photons\ cm^{-2}\ s^{-1}\ keV^{-1}$] at 1 keV. The fluxes and luminosities are absorption corrected.}
\end{center}
\end{table*}

The best fit to the $\rm 0.3 - 10\,keV$ {\em XMM-Newton} spectrum
of NGC 6240 is obtained using: (i)  three different plasma
temperatures; (ii) a highly absorbed power-law component; and
(iii) a neutral Fe K$\rm \alpha$ and K$\beta$ line (c.f. Table 1 and Fig. 7).
The broad-band fit with these spectral parameters results in an
acceptable fit ($\rm \chi_{r}^{2} = 0.93$ for 190 d.o.f.). In the
soft energy  band the spectrum is dominated by the cold ($\rm kT =
0.7\,keV$) and medium ($\rm kT = 1.4\,keV$) plasma components. The
medium component  is intrinsically absorbed with an absorbing
column of $(\rm 3.6 \pm 0.3) \cdot 10^{21}\ cm^{-2}$. For the $\rm
\alpha$ elements we find element abundances of (0.15 $\rm \pm$
0.08) relative to solar values. The Fe and  nickel abundance is
lower compared to that of the $\rm \alpha$ elements with ($0.05
\pm 0.01$); however, 
not significantly different.
The cooler plasma  component has an intrinsic column density of
 $\rm (2.0 \pm 0.3) \cdot 10^{21}\ cm^{-2}$, and a 
temperature of $\rm kT = (0.66 \pm 0.03)\,keV$. The available photon statistics does not allow us
to constrain the element abundances for the cooler plasma
component; therefore, we have fixed the $\rm \alpha$ and Fe and
nickel abundances to that of the hotter plasma component, which
results in a statistically acceptable fit for the soft energy
region. However, this should be handled with caution. Finally, including a third hot plasma component and an
absorbed power-law component allows us to explain the origin of the
ionized, Fe XXV and Fe XXVI line emission which were discussed in  detail in Section 3.2.2. This hottest plasma
component has a temperature of $\rm kT = (5.5 \pm 1.5)\,keV$
and an intrinsic absorption of $\rm (4.1 \pm 1.3) \cdot 10^{22}\
cm^{-2}$.
The APEC components in this model are regarded as a description for starburst activity.

A heavily absorbed power-law is required by the {\em XMM-Newton}
data. If neglected, significant residua remain above about 7 keV
and the fit is not statistically acceptable.
 We can confirm the {\em BeppoSAX} power-law with the {\em XMM-Newton}
data at a 4 $\rm \sigma$ significance level (according to an
F-test). The absorbing column density of $\rm (1.0 \pm 0.3)  \cdot
10^{24}\ cm^{-2}$ is a factor of two smaller than that found by
{\em BeppoSAX}. We believe that the photon index can be better
measured by {\em BeppoSAX} given the larger energy range over
which it was determined. Therefore, we have fixed the power-law
photon index to 1.8.  The absorbed power-law is indicative of the direct emission from the accretion disk corona.

We also note
that with the present statistics the fit is unable to
independently determine the temperature of the hot plasma
component, and the power-law photon index, and the column density.

\begin{figure}
\includegraphics[width=6.0cm,angle=90,clip=]{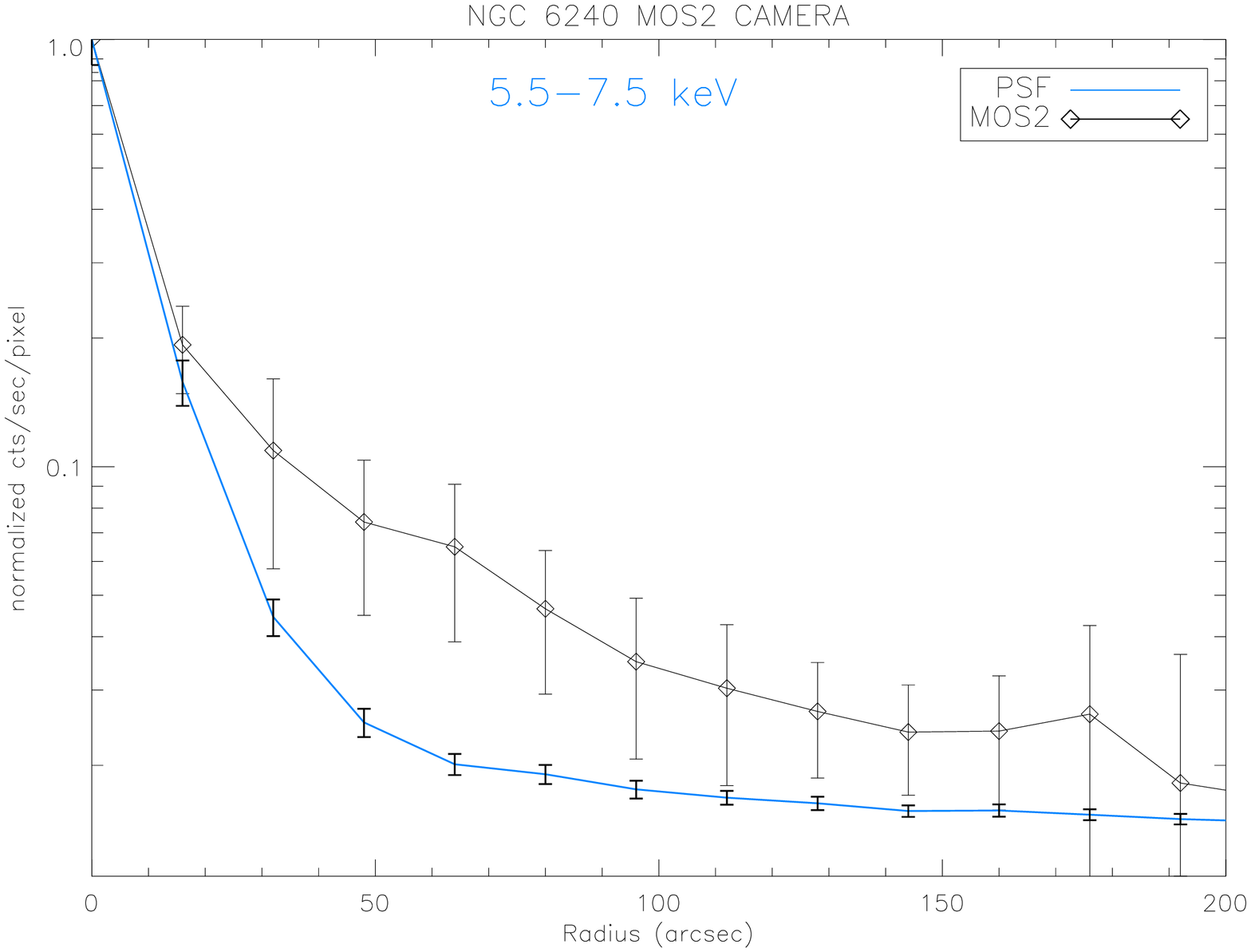}
\caption{\label{radprofile} The radial intensity profile of the
EPIC MOS1/2 data compared to the instrumental point spread function (PSF; normalized to the maximum flux). The solid line shows the MOS PSF of 6.5 keV and the photons have been selected in the 5.5-7.5 keV range to
match the emission from the Fe line complex. Extended emission
is detected at distances of about $30''$ from the X-ray
centroid position.}
\end{figure}


In order to cross--check the EPIC MOS results obtained from the (v)APEC
modelling with a different plasma code we substituted the three components
with the corresponding MEKAL model: in the soft regime the two vMEKAL
models and in the hard part of the spectrum a MEKAL model.
We find no significant difference in the absorbing column
densities and plasma temperatures with respect to the values listed in
Table 1. For the $\rm N_H$ values we get ($\rm 0.15 \pm 0.02$; 
$0.5 \pm 0.3$; and $\rm 3.5 \pm 1.1)$ $\rm \cdot 10^{22}$ $\rm
cm^{-2}$. The corresponding plasma temperatures are kT = ($\rm 0.62 \pm
0.03)\,keV$, ($\rm 1.2 \pm 0.4)\,keV$ and ($\rm 6.0 \pm 2.6)\,keV$.

A second test was performed to compare the results of the combined EPIC MOS data with the data of EPIC pn.
The resulting values $\rm N_H =
(0.19 \pm 0.03) \cdot 10^{22}$ atoms $\rm cm^{-2}$ with $\rm kT = (0.58 \pm
0.03)\,keV$, and $\rm N_H = (0.4 \pm 0.1) \cdot 10^{22}$ atoms $\rm
cm^{-2}$ with $\rm kT = (1.2 \pm 0.3)\,keV$ for the two 
vMEKAL components are in good agreement to the values of the EPIC MOS
data. The fitting with a MEKAL model yields in the values $\rm N_H = (3.8 \pm
2.0) \cdot 10^{22}$ atoms $\rm cm^{-2}$ and $\rm kT = (5.0 \pm 1.9)\,keV$.

\subsection{Comparison with local starbursts}
A striking similarity in the plasma parameters and absorbing columns
is found between NGC~6240 and the prototype local starburst
galaxy NGC~253. Pietsch et al. (2001) have analyzed the nuclear emission of NGC~253 with {\em
XMM-Newton}. They found that the nuclear spectrum can be
modelled by a three temperature plasma with the higher temperatures
increasingly absorbed. The derived temperatures for the nuclear region
are 0.56, 0.92 and 6.3\,keV. The corresponding absorbing column
densities are (0.34, 1.78, 13.2)$\rm \cdot 10^{22}$ atoms $\rm
cm^{-2}$. These temperatures and the corresponding absorbing
column densities are remarkably similar to those found in NGC~6240. In addition, Pietsch et al. detected He-like Fe K emission
at 6.68 keV, similar to the detection in NGC~6240. The non-detection of
the weaker H-like line in NGC~253 might be due to a lower signal-to-noise
compared to NGC~6240. The authors attribute the presence of the high
temperature component to emission from type Ib and type IIa SNRs.
In addition, similar soft X-ray emission lines from the two cooler plasma
components are detected in NGC~253, e.g. emission from Mg XI, Mg XII,
Si XIII, Si XIV (c.f. Section 4.1 of Pietsch et al).
The presence of different plasma components in the prototype local starburst
galaxy NGC~253 support our spectral modelling, and further suggests that there may
be a common underlying physical mechanism, which appear to be dominated
by emission from type Ib and type IIa SNRs.

\section{Comparison with Chandra}

In the following we combine the high throughput of  {\em XMM-Newton} with the high spatial
resolution of {\em Chandra}. While {\em Chandra} has resolved the nuclear region into two nuclei and
diffuse emission components (Komossa et al. 2003), {\em XMM-Newton} allows for X-ray spectroscopy with high statistics.

The inner $30''$ remain unresolved with {\em XMM-Newton} as obtained from the PSF.
The PSF was created simulating a point source at the same position on the detector as NGC~6240, and at an energy centered at 6.5 keV, using the XMM science simulator software {\tt SciSim}\footnote{{\tt http://xmm.vilspa.esa.es/scisim/}}. The energy selected corresponds to
the mean energy of the Fe K line complex. Serendipitous sources (localized through the XMMSAS task {\tt emldetect}) and out of time events were removed from the data. Each photon in the profile was corrected for vignetting and divided by the exposure map, to remove chip gaps and hot pixels.
Since neither the simulated source nor NGC~6240 show any pile-up, the sources were normalized to the maximum flux. Using such a PSF will allow us
to further investigate the spectral components around the two nuclei detected with {\em Chandra}.
The data were accumulated in the $\rm (5.5--7.5)\,keV$ energy range for the combined MOS1/2 data.
The radial intensity profile of the emission, 
together with the PSF are shown in Figure \ref{radprofile}.
Due to the large
width of the PSF, the (PSF-uncorrected) extended emission
appears to start at a distance greater than $30''$ from the center.  The spectral analysis of the extended emission
was extended up to a distance of $70''$. Beyond this distance the number of
counts is too small to make a significant contribution to the statistics.

We have performed a spectral analysis of the inner $30''$
independent from the results obtained from the total emission
($\rm 0 - 70''$). Interestingly, the spectral fit to the inner
$30''$ is consistent with the spectral fit components and
parameters obtained for the total emission (c.f. Fig. 7) except
for the flux emitted by the cold APEC component. Ninety-eight per
cent of the ($\rm 0.3 - 10\,keV$) flux is emitted in the inner $30''$ core. The derived
temperatures for the plasma components confirm the existence of
different plasma contributions in the core region.

In the following we compare the {\em XMM-Newton} plasma components
with the energy bands used in the high resolution {\em Chandra} image
 (Fig. 3 of Komossa et al.
2003). In the softest energy band (red coloured; $\rm 0.5-1.5\, keV$) the
emission from the 0.7 keV plasma component dominates the spectral
energy distribution, with a contamination level of about 10 per
cent from the 1.4 keV component. In the $\rm (1.5 - 5)\,keV$ band
(yellow and green coloured), we have a mixture of all three plasma
components and cannot uniquely attribute a single temperature to
the spatial location. Above 5 keV the hot 5.5 keV plasma component
dominates the spectrum and is responsible for the emission from Fe XXV and
Fe XXVI.
 This is confirmed by the {\em Chandra} observation;
 Komossa et al. (2003)  note that the neutral Fe K
line dominates the spectral energy distribution between (6--7) keV,
and noticed that some residua remain in their spectra which can be
attributed to ionized Fe K  emission.
In
addition,  the {\em XMM-Newton} data reveal a strong column
density gradient for the different plasma components for the inner
region, decreasing outwards from about $\rm 4.1 \cdot 10^{22}\,cm ^{-2}$ ($\rm kT =
5.5\,keV$), to $\rm 0.4 \cdot 10^{22}\,cm^{-2}$ ($\rm kT = 1.4\,keV$), and down
to $\rm 0.2 \cdot 10^{22}\,cm^{-2}$ ($\rm kT = 0.7\,keV$).


\begin{figure*}
\centerline{
\includegraphics[width=5.6cm,angle=0,clip=]{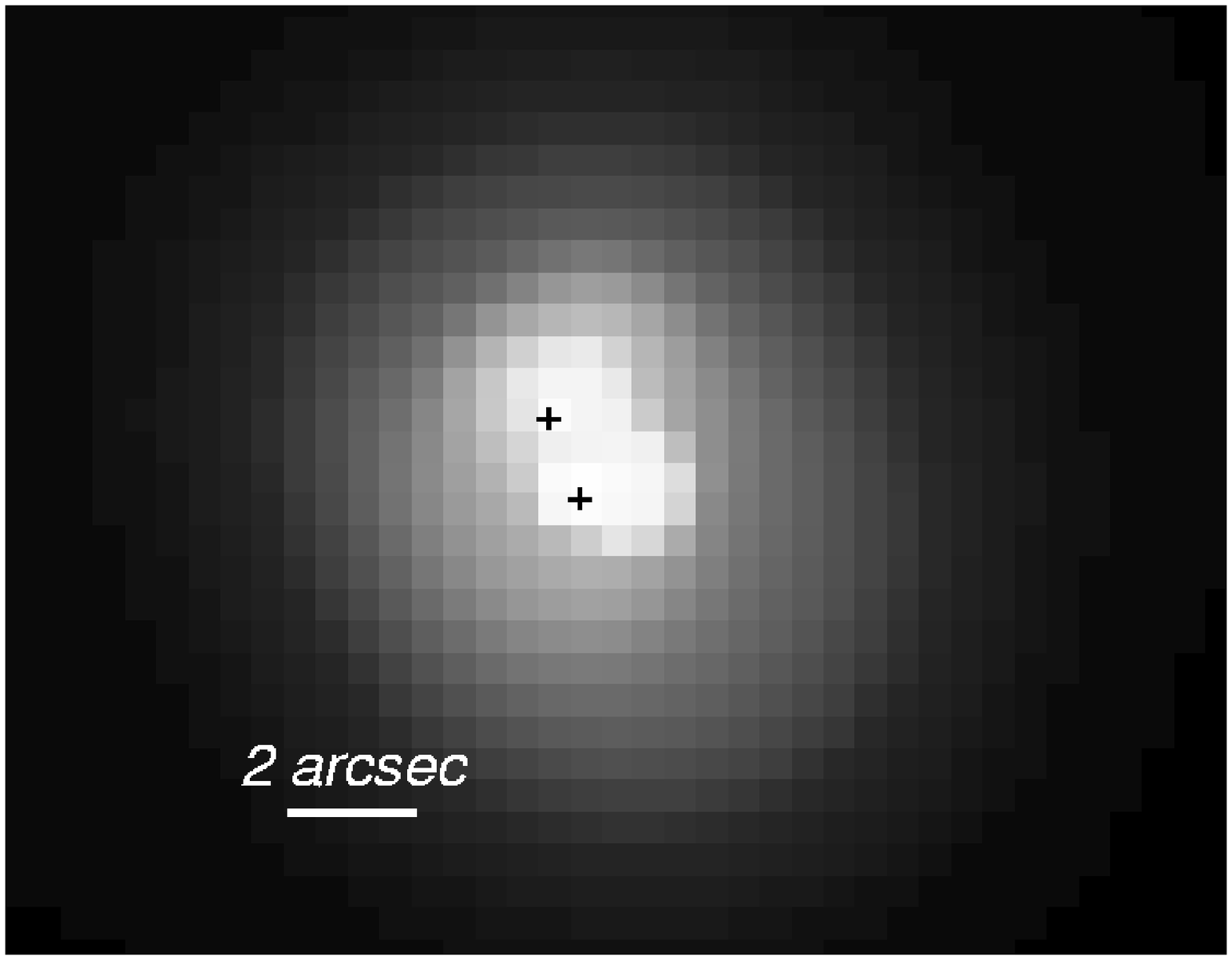}
\nolinebreak
\includegraphics[width=5.6cm,angle=0,clip=]{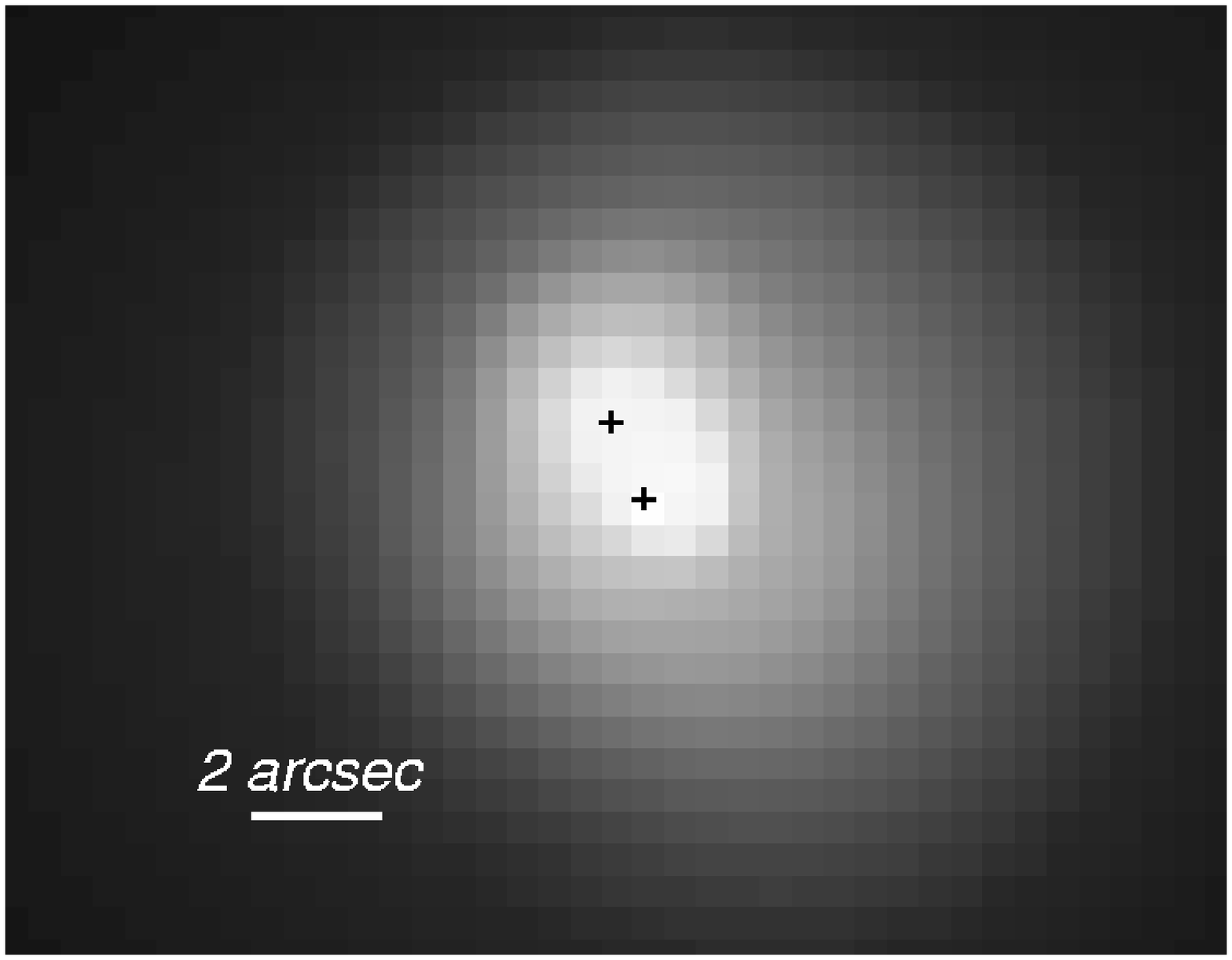}
\nolinebreak
\includegraphics[width=5.6cm,angle=0,clip=]{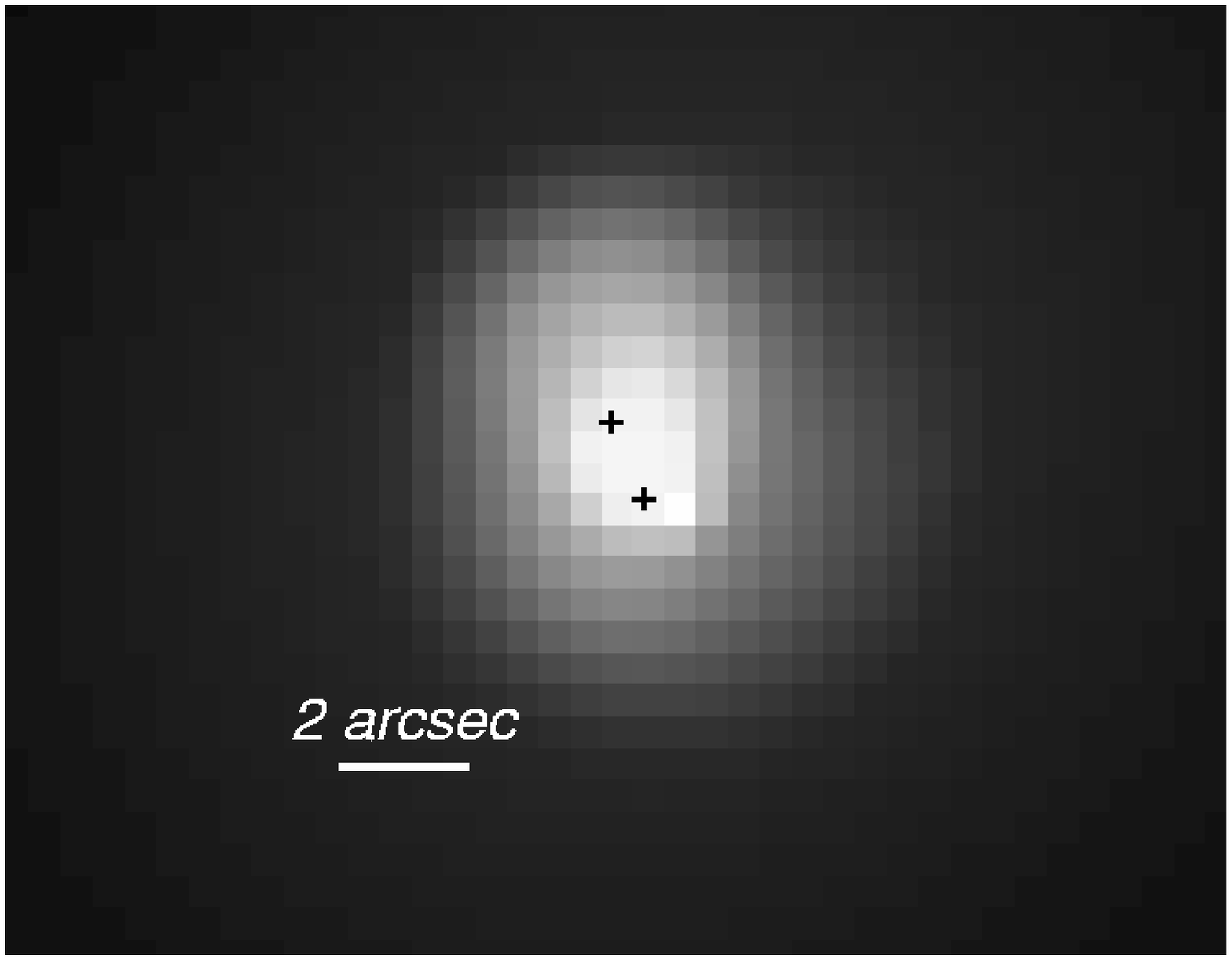}
}
\caption{
We have binned the {\em Chandra} data in three different energy bands as seen
by the XMM-Newton spectrum. The crosses mark the position of the two nuclei as detected with {\em Chandra}.
The left panel shows emission from the neutral Fe K line (selected
between 6.1 and 6.35 keV in the observers frame,
emission from the ionized lines (middle panel, energy range between 6.4 and 7 keV) as well as emission between 7  and 10 keV from the
highly absorbed power-law component (right panel). The spectral capabilities of {\em XMM-Newton} in combination with the
high spatial resolution of the {\em Chandra} telescope gives us a
more precise understanding of the nuclear region in
NGC 6240.}
\end{figure*}


\section{Summary}

{\em XMM-Newton} observations of the ULIRG NGC 6240 reveal:
(1)  For
     the first time, the presence of three distinct Fe K lines located
    at (6.41 $\pm$ 0.02)\,keV, (6.68 $\pm$ 0.02)\,keV and  (7.01
     $\pm$ 0.04)\,keV (in the rest frame of the source);

(2) That the broad-band spectrum is successfully described by the emission
from three different plasma components with temperatures of
 (0.66 $\rm \pm$ 0.03) keV,
 (1.4 $\rm \pm$ 0.2) keV, and
 (5.5 $\rm \pm$ 1.5) keV. 
The ionized lines can be explained by emission from the highest temperature
plasma components resulting in emission from Fe XXV and Fe XXVI.
The three plasma components are found at different locations
in the {\em Chandra} image, showing that the temperature and the column
density are increasing towards the center of NGC 6240;

 (3) A strong column density gradient for the different plasma components
 increasing with temperature from 
 (0.20   $\pm$ 0.03) $\cdot  {10^{22}\rm \ cm^{-2}}$, to
 (0.40   $\pm$ 0.03) $\cdot {10^{22}\rm \ cm^{-2}}$, to
 (4.1  $\pm$ 1.3) $\cdot {10^{22}\rm \ cm^{-2}}$;

(4) The presence of several emission lines in the soft energy band which is in agreement
     with previous ASCA measurements;

(5) A highly absorbed direct power-law component, confirming the BeppoSAX results;

(6) A striking similarity to the local starburst
    galaxy NGC 253.  In both, NGC 253 and NGC 6240,
     three plasma components are required, and their temperatures and
     column density gradients are remarkably similar; hence suggesting a similar underlying physical
     process at work in both galaxies.

The spectral capabilities of {\em XMM-Newton} in combination with the
high spatial resolution of the {\em Chandra} telescope gives us a
more precise understanding of the nuclear region in
NGC 6240 (c.f. Fig. 9).

\begin{acknowledgements}
Based on observations obtained with {\em XMM-Newton}, an ESA science mission with
instruments and contributions directly funded by ESA Member States and
the USA (NASA). We thank the anonymous referee for excellent suggestions and comments.
\end{acknowledgements}

\end{document}